\documentclass[preprint,prd,aps,showpacs,showkeys,nofootinbib]{revtex4-1}
\usepackage{amssymb}
\usepackage{amsmath}
\usepackage{graphicx}
\usepackage{subfigure}
\usepackage{float}
\usepackage{mathrsfs}
\usepackage{amsmath}

\begin{document}


\title{Primordial Black Hole Production in Inflationary Models of Supergravity with a Single Chiral Superfield }

\author{Tie-Jun Gao}
\email{tjgao@xidian.edu.cn}
\affiliation{School of Physics and Optoelectronic Engineering, Xidian University, Xi'an 710071, China}

\author{Zong-Kuan Guo}
\email{guozk@itp.ac.cn}
\affiliation{CAS Key Laboratory of Theoretical Physics, Institute of
Theoretical Physics, Chinese Academy of Sciences, P.O. Box 2735,
Beijing 100190, China}

\begin{abstract}
We propose a double inflection points inflationary model in supergravity with a single chiral superfield.  Such a model allows for the generation of primordial black holes(PBHs) at small scales, which can account for a significant fraction of dark matter. Moreover, in vacuum it is possible to give a small and adjustable SUSY breaking with a tiny cosmological constant.

\end{abstract}

\keywords{Inflation, Primordial black holes, Supergravity} \pacs{}

\maketitle

\section{Introduction\label{sec1}}
The nature of dark matter remains an open question in modern cosmology. One of the simplest explanations is assuming PBHs are a significant component of dark matter.  Recently, the detection of gravitational waves from binary black hole mergers by the LIGO and VIRGO opens a new window to probe black hole physics\cite{ref1,ref2,ref3} and to explore the nature of dark matter.

It is known that PBHs can arise from high peaks of the curvature power spectrum at small scales via gravitational collapse during the radiation era, which could constitute dark matter today. Such a peak in the power spectrum can be generated in a single-field inflation with an inflection point\cite{ref4} instead of using inflationary models with multiple fields\cite{ref5,ref6,ref7,ref8,ref9}. However, in \cite{ref10,ref11}, it is pointed out that close to the inflection point, the ultra-slow-roll trajectory supersede the slow-roll one, and thus, the slow-roll approximations cannot be used. So more precise approximation is used in\cite{ref12}, or the Mukhanov-Sasaki(MS) equation is numerically solved\cite{ref13}.
Recently, some inflationary models with an inflection point have been proposed to produce PBHs. For example, Ref.\cite{ref14} explores the possibility of forming PBHs in the critical Higgs inflation, where the near-inflection point is related to the critical value of the renormalization group equation running of both the Higgs self-coupling $\lambda$ and its non-minimal coupling to gravity $\xi$. And Ref.\cite{ref141} argues that diffusion could induce an enhancement of the power spectrum. In \cite{ref15} the author presents a toy model with a polynomial potential. PBHs product in Starobinsky's supergravity is presented in Ref.\cite{ref151} where the scalar belongs to the vector multiplet.
Ref.\cite{ref152} discusses the PBHs product in inflationary $\alpha-$attractors.
 Ref.\cite{ref16} provides a method to reconstruct the inflation potential from a given power spectrum, and gets a polynomial potential. In \cite{ref17} the authors present a single-field inflationary model in string theory which allows for the generation of PBHs.

Although cosmological inflation is now established by all precise observational data such as the WMAP~\cite{ref18} and Planck data~\cite{ref19}, the nature of inflation is still unknown. An interesting framework of inflation models building is to embed inflationary models into a more fundamental theory of quantum gravity, such as supergravity\cite{ref20,ref21,ref22}. However, in the supergravity based inflationary models, it always suffers from the so-called $\eta$ problem\cite{ref23}. The F-term of the potential is proportional to $e^{|\Phi|^2}$, which gives a contribution to the slow-roll parameter $\eta$ and breaks the slow-roll condition.
One way to overcome such obstacles is to invoke a shift symmetry of the K\"{a}hler potential, and add an extra chiral superfield, which can be stabilized at the origin during inflation\cite{ref24,ref25}. Another way is using a shift-symmetric quartic stabilization term in the K\"{a}hler potential instead of a stabilizer superfield \cite{ref26,ref27}, and SUSY tends to be broken at a scale comparable to the inflation scale in such kinds of models. For instance, by using logarithmic K\"{a}hler potential and cubic
superpotential, Ref.\cite{ref271} constructs an inflection point inflationary model which have a non-SUSY
de-Sitter vacuum responsible for the recent  acceleration of the Universe.
Whether SUSY is restored after inflation is quite important and worth discussing.  Ref.\cite{ref28} point out if SUSY breaks at a scale higher than the intermediate scale, the electroweak vacuum may be unstable, in addition, the related paper \cite{ref281} investigate this issue in detail, and find that the high-scale SUSY is still compatible with the known Higgs mass, though in a rather limited part of the parameter space.
Therefore, the authors investigate the SUSY breaking properties of the model in Ref.\cite{ref27} and study the conditions to restore SUSY after inflation.

In this paper, we shall consider the possibility to construct an inflection point inflationary model in supergravity with a single chiral superfield and focus on a superpotential with a sum of exponentials. We first assume SUSY restores after inflation, by fine-tuning the parameters of the model, such a superpotential can give a scalar potential with double  inflection points. One of the inflection points can make the prediction of scalar spectral index and tensor-to-scalar ratio consistent with the current CMB constraints at large scales. The other inflection point can generate a large peak of the power spectrum at small scales to arise PBHs. After inflation, one can obtain small SUSY breaking and a tiny cosmological constant by introducing a nilpotent superfield $S$\cite{ref27,ref29,ref30}.

The paper is organized as follows. In the next section, we setup the inflection point inflationary model in supergravity with SUSY restoration after inflation. In Section 3, we investigate the inflaton dynamics and compute the spectrum of primordial curvature perturbations of the model. In Section 4, we describe the mechanism of PBHs generation and calculate the mass distribution and abundance of PBHs. In Section 5, we consider the possibility to get a small cosmological constant with SUSY breaking. The last section is devoted to summary.

\section{inflection point inflation in supergravity with SUSY restored after inflation \label{sec2}}
In this section, we setup the inflection point inflationary model in supergravity and assumed that SUSY restores after inflation. The issue of SUSY breaking in vacuum is discussed in section 5.

Following Ref.\cite{ref27}, we consider a shift-symmetric K\"{a}hler potential of the form
\begin{eqnarray}
&&K=ic (\Phi-\bar{\Phi})-\frac{1}{2}(\Phi-\bar{\Phi})^2-\frac{\zeta}{4}(\Phi-\bar{\Phi})^4,
\label{kp6}
\end{eqnarray}
with $c$ and $\zeta$ are real constants. The real component $\phi$ of the chiral superfield $\Phi=(\phi+i\chi)/\sqrt{2}$ is taken to be the
inflaton and the quartic term serves to stabilize the field $\phi$ during inflation at $\langle\chi\rangle \simeq 0$ by making $\zeta$ sufficient large.

The scalar potential is determined by a given superpotential $W$ as well as K\"{a}hler potential,
which is given by
\begin{eqnarray}
&&V=e^{K/M_P^2}\Big[D_{\Phi_i}W(K^{-1})^{ij^*}D_{\Phi_j^*}W^*-3M_P^{-2}|W|^2\Big],
\label{infp}
\end{eqnarray}
where
\begin{eqnarray}
&&D_{\Phi}W=\partial_{\Phi}W+M_P^{-2}(\partial_{\Phi}K)W,
\end{eqnarray}
and $(K^{-1})^{ij^*}$ is the inverse of the K\"{a}hler metric
\begin{eqnarray}
&&K^{ij^*}=\frac{\partial^2K}{\partial\Phi_i\partial\Phi^*_{j}}.
\end{eqnarray}

In some inflationary models favoured by the CMB data, the scalar potential can be generated by a superpotential which
can be expanded as\cite{ref27,ref301},
\begin{eqnarray}
&&W=\sum_{n\geq0}a_n e^{-b_n\Phi},
\end{eqnarray}
where $a_n$ and $b_n$ are constants. If we require the SUSY preservation in vacuum with a vanishing cosmological constant, the F-term should be vanished $D_{\Phi}W=0$, and $V=0$ at $\Phi=0$, which requires the constraint
\begin{eqnarray}
&&W=\partial_{\Phi}W=0.
\end{eqnarray}


In order to produce a significant fraction of PBHs from primordial density perturbations which are consistent with the CMB constraints, we consider a superpotential of the form
\begin{eqnarray}
&&W=a_0(1+a_1 e^{-b_1 \Phi }+a_2 e^{-b_2 \Phi }+a_3 e^{-b_3 \Phi }).
\label{infp}
\end{eqnarray}
Such kinds of superpotential with exponential functions with two terms have been studied in the so-called racetrack model\cite{ref301,ref31,ref32} and in other models\cite{ref27}.
By solving the constraint (6),we can eliminate two of the parameters $a_1$ and $a_2$ as
\begin{eqnarray}
&&a_1\to \frac{b_2+a_3 b_2-a_3 b_3}{b_1-b_2},a_2\to \frac{-b_1-a_3 b_1+a_3 b_3}{b_1-b_2}.
\end{eqnarray}
Substituting the superpotential (7) and K\"{a}hler potential (1) into (2), one can get the scalar potential $V(\phi)$. In order to make the scalar potential predicted the primordial spectra consistent with the CMB data, give enough e-folding numbers and produce a significant fraction of PBHs in an interesting window for dark matter, we fine-tune the model parameters, and find some range in parameter space. We list two examples of parameter sets in Tab.1, and the scalar potential $V(\phi)$ for the first parameter set is depicted in Fig. 1.

\begin{figure}\small

  \centering
   \includegraphics[width=4in]{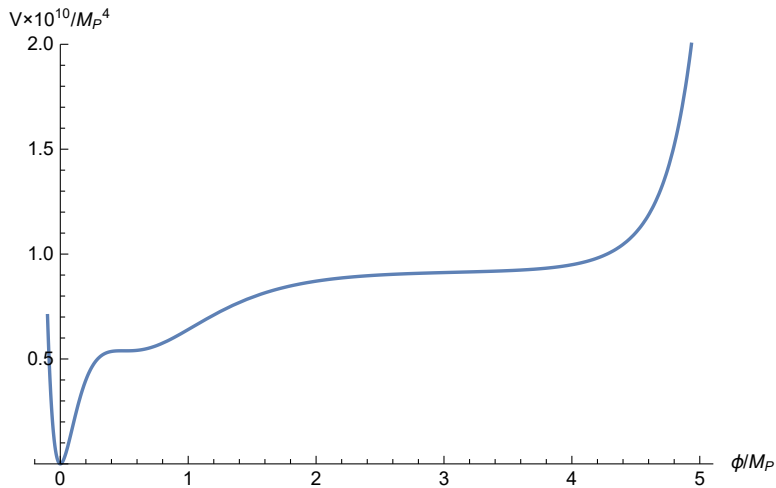}
     \caption{Scalar potential $V(\phi)$ for the first parameter set of Tab.1.}
    \label{fig1}
\end{figure}
We can see that the potential have two nearly inflection points, one of the inflection points can make the prediction of scalar spectral index and tensor-to-scalar ratio consistent  with the current CMB data, and the other one at small scales can generate a large peak in the power spectrum to arise PBHs. We will discuss these issues in the following sections.

\section{The spectrum of primordial curvature perturbations  \label{sec3}}

In the FRW homogeneous background, the Friedmann equation and the inflaton field equation can be written as
\begin{eqnarray}
&&H^2=\frac{1}{3M_P^2}\Big(\frac{1}{2}\dot{\phi}^2+V(\phi)\Big),
\end{eqnarray}
\begin{eqnarray}
&&\ddot{\phi}+3H\dot{\phi}+V'(\phi)=0,
\end{eqnarray}
where dots represent derivatives with respect to cosmic time and primes denote derivatives with respect to the field $\phi$.
The e-folding numbers from an initial time $t_i$ is defined as
\begin{eqnarray}
&&N_e(t)=\int_{t_i}^tH(t)dt,
\end{eqnarray}
where the e-folding number between the crossing time at the scale of $k_{0.001}$ and the time of the inflation end is required  in the range $50-60$. In this paper, we use $\Delta N_e^*$ to denote the e-folding number between $k_*=k_{0.05}$ and the end of inflation, which should be in the range $45-55$.

In the single-field slow-roll framework, the slow-roll parameters $\epsilon$ and $\eta$ can be calculated as
\begin{eqnarray}
&&\epsilon=\frac{M_P^2}{2}\Big(\frac{V'}{V}\Big)^2,\nonumber\\
&& \eta =M_P^2\Big(\frac{V''}{V}\Big).
\end{eqnarray}
However, in Ref.\cite{ref10,ref11} it is pointed out that close to the inflection point, the ultra-slow-roll trajectory supersedes the slow-roll one and thus one should use the slow-roll parameters defined by the Hubble parameter\cite{ref3201,ref3202,ref3203},
\begin{eqnarray}
&&\epsilon_H=-\frac{\dot{H}}{H^2},\nonumber\\
&&\eta_H=-\frac{\ddot{H}}{2H\dot{H}}=\epsilon_H-\frac{1}{2}\frac{d\ln\epsilon_H}{dN_e},\nonumber\\
&&\xi_H=\frac{\dddot{H}}{2H^2\dot{H}}-2\eta_H^2=\epsilon_H \eta_H-\frac{d\eta_H}{dN_e}.
\end{eqnarray}
Since near the inflection point, the potential becomes extremely flat, so the slope term $V'(\phi)$ of Eq.(10) may reduce drastically, which means $|\eta_H|=|-{\ddot{\phi}}/{H\dot{\phi}}|\simeq3$. For a nearly inflection point there exist a range satisfies $\ddot{\phi}+3H\dot{\phi}=-V'(\phi)>0$ and $\dot{\phi}<0$, which leads to $|\eta_H|>3$, thus the slow-roll approximation is no longer applicable.
The Hubble slow-roll parameters $\epsilon_H$ and $\eta_H$  as functions of the e-folding number $\Delta N_e^*$  for parameter set 1 are show in Fig.2

\begin{figure}\small

  \centering
   \includegraphics[width=4in]{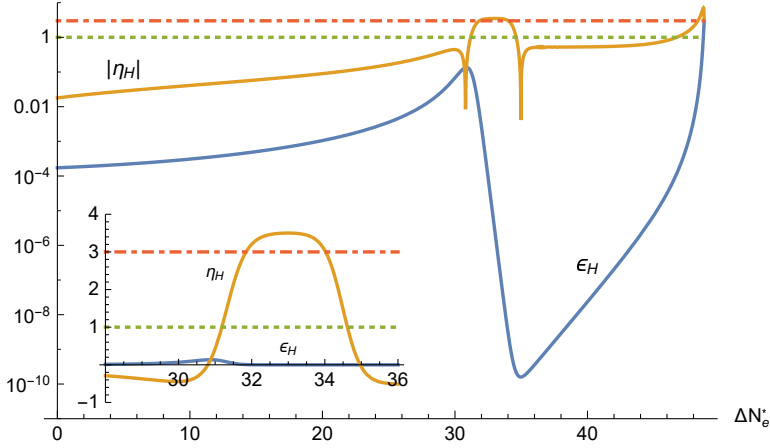}
     \caption{Slow-roll parameters $\epsilon_H$(blue solid line) and $\eta_H$(orange solid line) as functions of the e-folding number $\Delta N_e^*$ for parameter set 1. The green dashed line and red dot-dash line indicate the values 1 and 3 respectively. The inset zooms into the relevant range of $\Delta N_e^*$ for the ultra-slow-roll inflation where the slow-roll approximation  fails.}
    \label{fig1}
\end{figure}
We can see that near the inflection point the slow-roll parameter $|\eta_H|>3$, so the slow-roll approximation fails. There is a valley on the curve of $\epsilon_H$, which can give rise to a large peak of the primordial power spectrum.

At the leading order, the scalar spectral index and its running as well as the tensor-to-scalar ratio can be expressed using $\epsilon_H$, $\eta_H$ and $\xi_H$  as
\begin{eqnarray}
&&n_s=1-4\epsilon_H+2\eta_H,\nonumber\\
&&\alpha=\frac{dn_s}{d\ln k}=10\epsilon_H \eta_H-8 \epsilon_H^2-2 \xi_H, \nonumber\\
&&r=16\epsilon_H.
\end{eqnarray}
Then the scalar power spectrum can be approximately calculated using the expression
\begin{eqnarray}
&&P_R\simeq\frac{1}{8\pi^2M_P^2}\frac{H^2}{\epsilon_H}.
\end{eqnarray}

The observational constraints from Planck on the scalar spectral index $n_s$ and its running $\alpha$, the tensor-to-scalar ratio $r$ and amplitude of the primordial curvature perturbations $A_s$ at $68\%$ C.L. at a scale $k_*=0.05\mathrm{Mpc}^{-1}$ are \cite{ref19}
\begin{eqnarray}
&&n_s=0.9650\pm0.0050,\nonumber\\
&&\alpha=-0.009\pm0.008,\nonumber\\
&&r<0.07,\nonumber\\
&&A_s=2.2\pm0.1\times10^{-9}.
\end{eqnarray}

However, in order to compute the power spectrum near the inflection point more reliably, one must solve the MS equation of mode function
\begin{eqnarray}
&&\frac{d^2u_k}{d\tau^2}+\Big(k^2-\frac{1}{z}\frac{d^2z}{d\tau^2}\Big)u_k=0,
\end{eqnarray}
where $\tau$ denotes conformal time and $z\equiv\frac{a}{\mathcal{H}}\frac{d\phi}{d\tau}$.
The initial condition for Eq.(17) is taken to be the Bunch-Davies type\cite{ref33}
\begin{eqnarray}
&&u_k\rightarrow \frac{e^{-ik\tau}}{\sqrt{2k}},\;\;  \text{as}\;\; \frac{k}{aH}\rightarrow\infty.
\end{eqnarray}
For the purpose of numerical simulation, the MS equation can be written in terms of $N_e$ as the time variable\cite{ref13}
\begin{eqnarray}
&&\frac{d^2u_k}{dN_e^2}+(1-\epsilon_H)\frac{du_k}{dN_e}+[\frac{k^2}{\mathcal{H}^2}+(1+\epsilon_H-\eta_H)(\eta_H-2)-\frac{d\epsilon_H-\eta_H}{dN_e}]=0,
\end{eqnarray}
and the power spectrum can be calculated by
\begin{eqnarray}
&&\mathcal{P}_\mathcal{R}=\frac{k^3}{2\pi^2}\Big|\frac{u_k}{z}\Big|^2_{k\ll\mathcal{H}}.
\end{eqnarray}

In Fig.3, we plot the primordial power spectrum as a function of the e-folding number $\Delta N_e^*$ with parameter set 1. The solid line is calculated from the solutions to the MS equation, while the dashed line is calculated by using the approximation(15), which underestimates the power spectrum, thus couldn't be used to obtain the PBH abundance and mass.  We can see that there is a large peak at small scales, with a height of about seven orders of magnitude more than the spectrum at CMB scales, which can generate PBHs via gravitational collapse.
\begin{figure}\small

  \centering
   \includegraphics[width=4in]{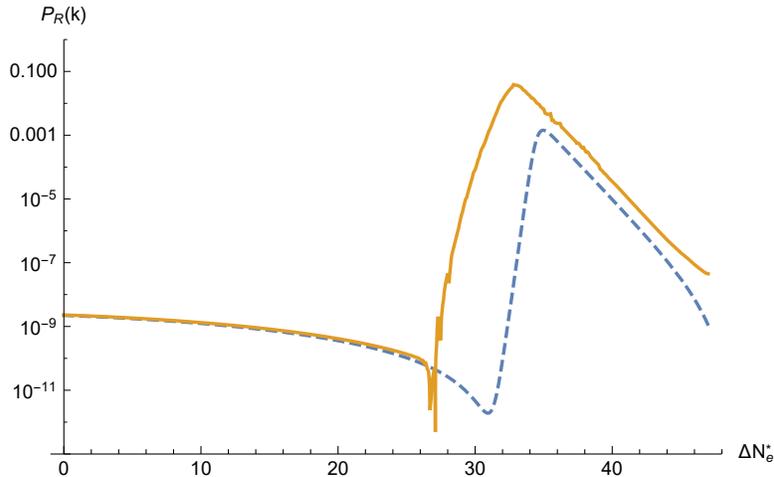}
     \caption{Primordial power spectrum as a function of the e-folding numbers $\Delta N_e^*$ for parameter set 1. The blue solid line is calculated by using the MS equation and the orange dashed line is  calculated by using the approximation(15).}
    \label{fig1}
\end{figure}

Our numerical results of inflationary dynamics corresponding to the two parameter sets are presented in Tab.2, and they are in agreement with the  current  CMB constraints on the primordial spectra(16).
\begin{table}

\begin{tabular}{c||c|c|c|c|c|c}
&$a_0$&$a_3$&$b_1$&$b_2$&$b_3$&$c$\\
\hline  
1&$4.35\times10^{-6}$&$7\times10^{-8}$&$3.05$&$6.3868164$&$-4.4$&$2.8$\\

2&$4.06\times10^{-6}$&$1\times10^{-6}$&$2.89$&$7.251197$&$-3.2$&$2.85$\\


\end{tabular}
\caption{Examples of parameter choice }
\end{table}

\section{Production of primordial black holes \label{sec3}}
The mechanisms of PBHs production have been studies in several references \cite{ref340,ref341,ref342,ref343,ref344}.
And in this section, we will calculate the PBH abundance using the Press-Schechter approach\cite{ref34} of gravitational collapse.
When a large amplitude of primordial fluctuations, which is generated at small scales during inflation and re-enters the Hubble horizon after
inflation. It undergoes gravitational collapse and form PBHs if the density fluctuation of matter is significantly large.

The mass of the resulting PBHs is assumed to be directly proportional to the horizon mass at re-entry time,
\begin{eqnarray}
&&M=\gamma M_H=\gamma\frac{4}{3}\pi \rho H^{-3}.
\end{eqnarray}
It can be approximated as\cite{ref13}

\begin{eqnarray}
&&M\simeq10^{18} \text{g} \Big(\frac{\gamma}{0.2}\Big)\Big(\frac{g_*}{106.75}\Big)^{-1/6}\Big(\frac{k}{7\times10^{13}\text{Mpc}^{-1}}\Big)^{-2},
\end{eqnarray}
where $\gamma\sim 0.2$ is a proportionality constant, which depends on the details of the gravitational collapse\cite{ref35}, and $g_*\sim 106.75$ is the effective degrees of freedom for energy density, which is equal to that of entropy density.

In the context of the Press-Schechter model of gravitational collapse, assuming that the probability distribution of density perturbations is Gaussian with width $\sigma(M)$, the formation rate of PBHs, which we denote as $\beta(M)$ is given by
\begin{eqnarray}
\beta(M)\equiv\frac{\rho_{PBH}(M)}{\rho_{tot}}&&=\frac{1}{\sqrt{2 \pi \sigma^2(M)}}\int^{\infty}_{\delta_c}d\delta \text{exp}\Big(\frac{-\delta^2}{2 \sigma^2(M)}\Big)\nonumber\\
&&=\frac{1}{2}\text{erfc}\Big(\frac{\delta_c}{\sqrt{2\sigma^2(M)}}\Big),
\end{eqnarray}
where $\delta_c\simeq0.45$\cite{ref351,ref352} denotes the threshold of density perturbations for the collapse into PBHs.
Here $\sigma^2(M)$ is the variance of the comoving density perturbations coarse grained at a scale $R=1/k$, during the radiation-dominated era, which is given by,
\begin{eqnarray}
&&\sigma^2(M(k))=\frac{16}{81}\int\frac{dq}{q}(qR)^4P_{R}(q)W(qR)^2,
\end{eqnarray}
where $P_{R}$ is the power spectrum of the primordial comoving curvature perturbations, and the smoothing window function $W(x)$ is taken to be a Gaussian $W(x)=exp(-x^2/2)$.
Then integrating over all masses $M$ one can get the  present abundance of PBHs,
\begin{eqnarray}
&&\Omega_{PBH}=\int\frac{dM}{M}\Omega_{PBH}(M),
\end{eqnarray}
with
\begin{eqnarray}
&&\frac{\Omega_{PBH}(M)}{\Omega_{DM}}=\Big(\frac{\beta(M)}{1.6\times10^{-16}}\Big)\Big(\frac{\gamma}{0.2}\Big)^{3/2}\Big(\frac{g_*}{106.75}\Big)^{-1/4}\Big(\frac{M}{10^{18} \text{g} }\Big)^{-1/2},
\end{eqnarray}
where $\Omega_{DM}\simeq0.26$ is the total dark matter abundance\cite{ref36}.

In Fig.4, we plot the factional abundance of PBHs for the first parameter set in Tab.1 and the observational constraints from Ref.\cite{ref37,ref371}
\begin{figure}\small

  \centering
   \includegraphics[width=4in]{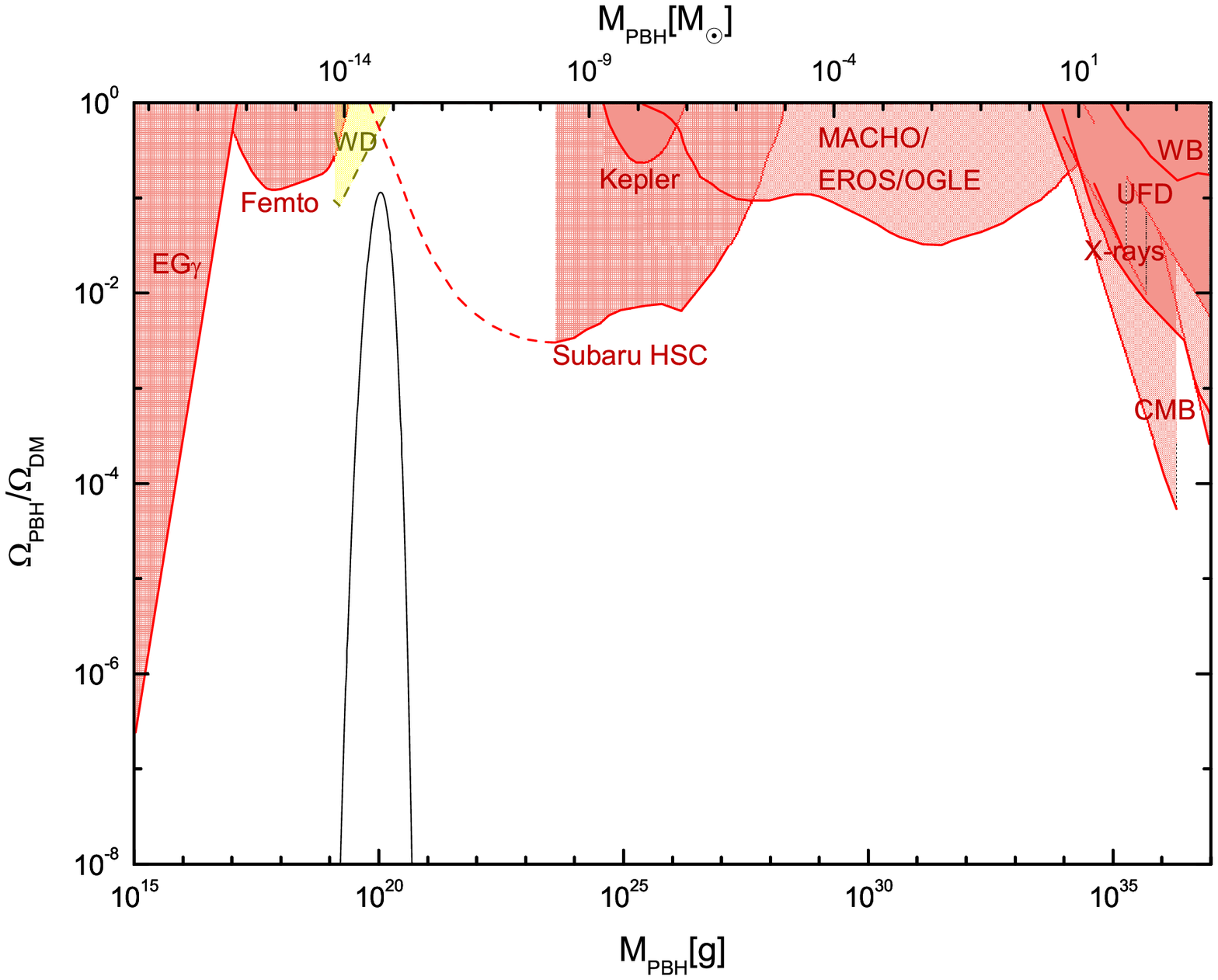}
     \caption{The factional abundance of PBHs for the first parameters set in Tab.1 and the observational constraints
      including the extra-galactic gamma-rays from the PBH evaporation (EG$\gamma$)\cite{ref50},
      the femtolensing of gamma-ray bursts(Femto)\cite{ref51}, the white dwarfs explosion (WD)\cite{ref52}, the microlensing events with Subaru HSC (Subaru HSC)\cite{ref53}(The
red dotted line shows the uncertain constraint of HSC),
with Kepler satellite (Kepler)\cite{ref54}, with MACHO/EROS/OGLE\cite{ref55,ref56,ref57},
the CMB measurements\cite{ref58,ref59}, the radio and X-ray observations \cite{ref60,ref61}, the
dynamical heating of dwarf galaxies and ultra-faint dwarf galaxies
(UFD) \cite{ref62,ref63} and the distribution of wide binaries (WB)\cite{ref64}  }
    \label{fig1}
\end{figure}
The numerical results of the PBHs mass and abundance for the two parameters sets are listed in Tab.2.
\begin{table}

\begin{tabular}{c||c|c|c|c|c|c|c}
&$n_s$&$r$&$\alpha$&$\Delta N_e^*$&$P_R^{peak}$&$M_{PBHs}^{peak}/M_{\odot}$&$\Omega_{PBH}/\Omega_{DM}$\\
\hline  
1&$0.9635$&$0.00276$&$-0.00369$&$48.66$&$0.0369$&$5.50\times10^{-14}$&$0.114$\\

2&$0.9649$&$0.00257$&$-0.00286$&$46.29$&$0.0032$&$1.39\times10^{-16}$&$0.022$\\


\end{tabular}
\caption{Results of inflationary observables at CMB and small scales.}
\end{table}

\section{SUSY breaking and vacuum structure \label{sec4}}

After inflation, in order to obtain small SUSY breaking and a very small cosmological constant, it is possible to extend the K\"{a}hler potential and superpotential with a nilpotent superfield $S$\cite{ref29,ref30}. Since such a method is discussed in Ref.\cite{ref27}, we just list some conclusions.

The K\"{a}hler potential and superpotential can be extended as
\begin{eqnarray}
&&K=K^{inf}+S\overline{S},\nonumber\\
&&W=W^{inf}+W_0+\mu^2S,
\end{eqnarray}
where $W_0$ and $\mu$ are constants and the upper indexes denote the quantities in the inflation sector as in Eqs.(1) and (7).
In the leading order of $|W_0|$ and $|\mu|^2$, the vacuum expectation value of the inflation is
\begin{eqnarray}
&&\langle\Phi\rangle=-\frac{W_0K_{\Phi}^{inf}}{W^{inf}_{\Phi\Phi}}.
\end{eqnarray}
Then the vacuum energy becomes
\begin{eqnarray}
&&V=|\mu|^4-3|W_0|^2.
\end{eqnarray}
Since the SUSY breaking scale $|D_SW|=|\mu|^2$ can be chosen freely, by fine-tuning the parameters one can get a small and adjustable cosmological constant of order $10^{-120}$ responsible for the current accelerated expansion of the Universe.

The role of the nilpotent superfield in this work just leads to a tiny cosmological constant, and cannot be used for large resolution of masses between known particles and their superpartners in particles physics.
In addition, nilpotent superfields and related non-linear realizations are not necessary for SUSY breaking. Other models having linearly realizations and spontaneously SUSY breaking are discussed in \cite{ref65} with a single vector (inflaton) superfield instead of a single chiral inflaton superfield.

\section{Summary \label{sec5}}

In this paper, we propose a double inflection point inflationary model in supergravity with a single chiral superfield. We focus on a superpotential with a sum of exponentials and assume the SUSY restores after inflation. We find that such a superpotential can give a scalar potential with double inflection points. We investigate the inflaton dynamics and compute the spectrum of primordial curvature perturbations, and find that the inflection point at large scales can make the predicted of scalar spectral index and tensor-to-scalar ratio consistent with the current CMB data.

The other inflection point at small scales can generate a large peak in the power spectrum with a height of about seven orders of magnitude more than the spectrum at CMB scales. When the significantly large amplitude of primordial fluctuation re-enters the Hubble horizon after inflation, it will undergo gravitational collapse and form PBHs. Moreover, we describe the mechanism of PBHs generation, and get the mass and abundance of PBHs which can  account for a significant component of dark matter.

After inflation, it is possible to give a small and adjustable SUSY breaking and a tiny cosmological constant by extending the K\"{a}hler potential and superpotential with a nilpotent superfield. Since the SUSY breaking scale is much lower than the inflation scale, the effects of nilpotent superfield $S$ on the inflationary dynamics can be negligible.

\begin{acknowledgments}
This work was supported by "the National Natural Science Foundation of China" (NNSFC) with Grant No. 11705133, and "the Fundamental Research Funds for the Central Universities"  No.JBF180501. ZKG is supported in part by the National Natural Science Foundation of China Grants No. 11575272, No. 11690021 and No. 11335012.
\end{acknowledgments}

\end{document}